# Conscious Commerce – Digital Nudging and Sustainable E-commerce Purchase Decisions

**Full research paper**

**Milad Mirbabaie**
Department of Information Systems
Paderborn University
Paderborn, Germany
Email: milad.mirbabaie@uni-paderborn.de

**Julian Marx**
Department of Computer Science and Applied Cognitive Science
University of Duisburg-Essen
Duisburg, Germany
Email: julian.marx@uni-due.de

**Johanna Germies**
Department of Computer Science and Applied Cognitive Science
University of Duisburg-Essen
Duisburg, Germany
Email: johanna.germies@uni-due.de

## Abstract

So-called 'fast fashion' consumption, amplified through cost-effective e-commerce, constitutes a major factor negatively impacting climate change. A recently noted strategy to motivate consumers to more sustainable decisions is digital nudging. This paper explores the capability of digital nudging in the context of green fashion e-commerce. To do so, digital default and social norm nudges are tested in an experimental setting of green fashion purchases. An online experiment (n = 320) was conducted, simulating an online retail scenario. Results failed to show statistically significant relationships between nudging strategies and purchase decisions. However, explorative analyses show a backfiring effect for the combination of nudges and thus, reveal a hitherto neglected impact of participants' identification on the effectiveness of the digital nudging strategies. Consequently, this study contributes to digital nudging literature and informs practice with new insights on effective choice architectures in e-commerce.

# 1　Introduction

In the last years, climate change has become a severe threat to the global environment and human life (Fawzy et al. 2020). The fashion industry is the second largest contributor to environmental problems after the oil industry (Dhir et al. 2020). Many fashion consumers bear the belief that environmentally sustainable behaviour is important and state that they are willing to purchase green fashion products (GFP), that is ecologically produced fashion, for even higher prices (Haider et al. 2019; Moraes et al. 2012). However, a study shows that only 14% of consumers are actually buying fashion products that are made out of sustainable materials (Statista 2018).

Previous research investigated how to sustainably change attitudes and motivate environmentally conscious consumer behaviour (Gupta and Ogden 2006) and many researchers have already successfully applied *digital nudges* – defined as certain design changes made in a choice architecture to encourage decision-making processes and alter individuals' behaviour (Hummel and Maedche 2019; Kroll et al. 2019; Weinmann et al. 2016). Digital nudging functions without creating any restrictions or forcing behaviour but rather, for instance, easing decision-making or enhancing information transparency (Halpern 2015; Thaler and Sunstein 2009). Studies show that digital nudges seem to be particularly effective in steering online buying behaviour (Mols et al. 2015; Myers and Souza 2020). However, its applicability has largely focused on online retail of so-called low involvement products such as groceries (Demarque et al. 2015; Ingendahl et al. 2020) or furniture (Van Gestel et al. 2020) and research in the context of high-involvement products, like fashion products has been neglected so far. Findings of previous studies are not easily transferable to fashion products as purchasing behaviour of high and low involvement products differs due to linked factors like cognitive performance or emotional attachment (Huh et al. 2014). Although the fashion industry is aiming to reduce its environmental footprint, tackling climate change requires e-commerce choice architectures that default to "sustainability by design". Therefore, investigating the application of digital nudging in the fashion context is important. Based on this reasoning, the following research question will be examined in this paper:

***RQ:*** *How does digital nudging enhance environmentally sustainable consumer behaviour in the context of online fashion retail?*

To answer this research question, we propose a quantitative between-subject online experiment (n=320). The results of this study provide further insight into the intricate dynamics and relations of digital nudges in the context of GFP such as the relationships between social norms, default settings and environmentally friendly purchase decisions. The study further discusses the limits and potentials of applying nudging strategies in e-commerce settings as a highly relevant field to information systems (IS) research. Moreover, this study follows up on practical contributions of previous studies in that assumptions can be made about the extent to which nudges are applicable in the fashion online retail sector and provides practical recommendations for designing choice architectures that foster more environmentally sustainable behaviour while maintaining sales volume.

# 2　Related Work

## 2.1　Environmentally Sustainable Purchasing Decisions

In general, a decision-making process is shaped by two cognitive processes: First, the intuitive and automatic process (system 1) and secondly the reflective, rational process (system 2) (Halpern 2015; Thaler and Sunstein 2009). In some situations, humans think consciously about their actions – such as during processes like learning to drive a car – in other situations, however, decisions are more likely taken by rapid and instinctive thinking, for instance tying the shoes in the morning (Kahneman, 2011). However, human decision-making, and especially consumer decision-making is not always rational and predictable as it is always influenced by other determinants (Thaler and Sunstein 2009). There are certain predictable errors in human decision-making that must be considered as influences on decision-making processes. Four principle underlying factors have been identified to impact the consumer decision-making process: cultural influences, such as values and affiliation with subcultures or social classes; individual factors defined by gender, age of the person or self-concept; psychological factors such as individual knowledge, memory capacity, beliefs and attitudes; and finally social factors, defined by the influences of a reference group, opinion leaders or the family (Lamb et al. 2011).

Referring to cultural influences and social classes, in the context of environmental consciousness, higher-income status, educational and economic level, as well as a stronger identification with (Gupta and Ogden 2006)**.** Furthermore, existing research found that individuals who closely identified with representatives that support sustainable behaviour show an enhanced purchasing behaviour (Bly et al. 2015). Additionally, psychological factors such as when individuals are exposed to a mass of available information or a complex issue that requires expert knowledge that individuals do not naturally inhibit seems to impact decision-making processes (Halpern 2015; Mols et al. 2015; Mont et al. 2015). The influences described above often cause problems for individuals to translate beliefs into actions. Besides monetary incentives, bans, or developing laws that impose penalties when they are broken, a digital nudging has been identified as a potentially effective strategy to motivate consumer behaviour while sustaining their freedom of choice.

## 2.2 Motivating Decision-Making with Digital Nudging

The term *nudging* was introduced by Thaler and Sunstein (2009), who argued that by altering the choice architecture in which humans' decision-making is being performed, individuals can be subtly directed towards or give gentle hints for actions and desirable outcomes. Nudging, as an alternative to design policies such as bans or restrictions, subtly motivate individuals to perform certain behaviour with objectively positive consequences for them (Sunstein 2014). As a result of steadily increasing digitization and many decision-making processes shifting to the virtual realm, the concept of *digital nudging* has been introduced by Weinmann et al. (2016). It describes the use of certain design elements in user interfaces to guide users through the digital choice environment and affect choices while using IS.

Digital nudging can be applied to predictively affect user behaviour leading to the possible selection of a for instance pre-defined choice but always leaving the user with the freedom to make a different choice (Schneider et al. 2018; Weinmann et al. 2016). This describes a dominant notion in nudging theory – *libertarian paternalism* – that free decision-making of humans needs to be ensured to reject that recommendation (Thaler and Sunstein 2009). Nudges, can be looked at as *"encouraging or guiding behaviour without mandating or instructing, and ideally without the need for heavy financial incentives or sanctions"* (Halpern, 2015, p.22). It has been shown that through its application, the gap between attitude and behaviour can be closed and people can be motivated to act in line with their beliefs (Van Gestel et al. 2020; Thaler and Sunstein 2009).

A constantly growing range of nudging strategies is being developed. However, two particular design types seem the most relevant to this study: studies show that digital nudges making a pre-selection for users (default nudges) or providing them with information about purchasing decision of other customers, which are determined as social norms (social norm nudges), seem to be particularly effective in steering online buying behaviour (Djurica and Figl 2017; Mols et al. 2015; Myers and Souza 2020). Pre-selection of preferred actions as default rules has been found to be the most effective strategy to enhance the motivation for certain actions and serve as a well-established strategy to motivate behaviour change (Hummel and Maedche 2019; Thaler and Sunstein 2009), since they make use of automated cognitive processes (Benartzi et al. 2017; Sunstein 2014; Whyte et al. 2012). It has been found that when people are automatically enrolled in default programs, they are more likely to follow through with the pre-selected option than to opt-out, as opting-out is an additional conscious effort on the individuals (Van Gestel et al. 2020; Schneider et al. 2018). In the following, relevant research that deals with digital nudging to motivate environmentally sustainable behaviour will be presented to derive the research hypotheses.

## 3 Hypotheses Development

The potential of digital nudges to purposefully direct individuals' decision-making in IS to meet the desired outcome such as promoting more environmentally conscious consumer behaviour has been recognized not only within business contexts (Bammert et al. 2020), but also in relation to an end user (Henkel et al. 2020). Prior studies in have already applied different default nudges to encourage environmentally conscious consumer behaviour (Hummel and Maedche 2019). Based on the existing research, it can be assumed that default nudges successfully motivate consumer behaviour in the context of fashion retail as well. Consequently, the following hypothesis can be derived:

***H1a:*** *The use of digital default nudges in online fashion retail enhances GFP choices compared to decision-settings in which no nudge is applied.*

Moreover, research on nudging builds on the notion that decision-making is often motivated by individuals comparing their own performance or beliefs with others (Claudy et al. 2013; Turner et al. 1979). Existing research has proven the high potential of social norm nudges in contexts relating to environmentally sustainable behaviour such as enhancing the willingness to reuse towels in hotels (Goldstein et al. 2008), reducing paper waste (Chakravarty and Mishra 2019) or promoting recycling behaviour (Czajkowski et al. 2019). Demarque et al., (2015) highlight that different tiers of social norm nudges have the same impact on behaviour change. The researchers applied strong (70%) and weak social norms (9%) to represent the percentage of the previous participants who have purchased at least one eco-friendly product. Therefore, it can be expected that low social norms can also motivate behaviour change in the context of online retail. Mirroring H1a, this leads to the following hypothesis:

**H1b:** *The use of digital social norm nudges in online fashion retail enhances GFP choices compared to decision settings in which no nudge is applied.*

In addition to findings that support the effectiveness of social norm and default nudges, previous studies have demonstrated that the combination of different nudging strategies can increase their effectiveness (Ingendahl et al. 2020; Kroll et al. 2019). It can therefore be assumed, that a combined application of default and social norm nudges in an online retail environment enhances an environmentally friendly behaviour, and that the number of selected environmentally sustainable products can be increased. This leads to the following hypothesis:

**H2:** *The use of digital default nudges and social norm nudges combined in online fashion retail leads to more GFP choices than with no or single nudge applied.*

The hypotheses are summarized in Figure 1, which depicts the proposed research model. Accordingly, the impact of the applied digital nudging strategies on product selection is investigated aiming for enhancing the selected number of GFPs.

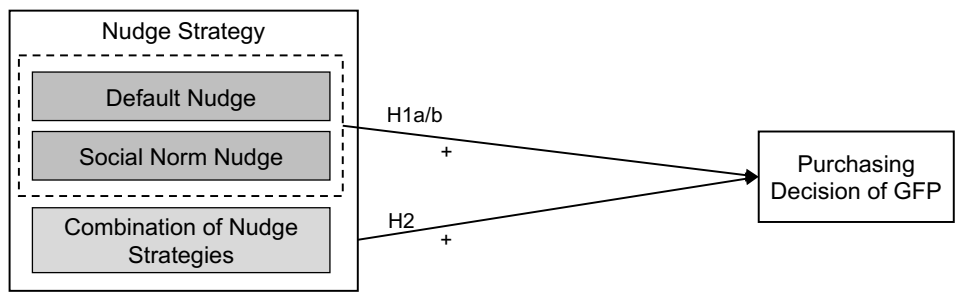

*Figure 1: Overview of the research model.*

## 4  Research Design

A four condition between-subject online experiment was conducted to imitate an e-commerce environment. The data collection took place in January 2021 using the online survey tool *LimeSurvey*. The experiment was conducted online and in German language. First, the participants were shown six product categories – jackets, pants, t-shirts, beanies, shoes, backpacks –, each with two product options and were asked to indicate which of the products they would rather buy. In the second part of the survey all participants were asked whether their decision felt autonomous and to what extent they identify with environmentally sustainable practices. Afterward, they were asked demographical questions, and lastly, they were presented with a debriefing that outlines the goal of the study. The inverted set-up ensures that the participants complete the task without being primed (McCambridge et al. 2012). The participants were recruited by face-to-face or through social media advertisements. The participants were allocated randomly in one of four treatment groups: *default nudge*, *social norm nudge*, *combination*, and *control*.

Each of the six product categories consisted of two different products: one sustainable option and one standard option. For this purpose, twelve suitable products had to be identified. The displayed options, although not identical, were similar in their optics to reduce possible biases. Moreover, as brand preferences are found to have an impact on decision-making, brand names or logos were not presented. Participants were given indicators such as a price and the material composition of the products. While the informative text of the standard product stated that the product consists of lower quality material (e.g.: "material**:** 70% polyester, 30% viscose"), the sustainable product was declared as a higher quality product with less environmental impact (e.g. "material: 100% cotton. Eco-friendly material with low $CO_2$ emissions"). Furthermore, the displayed informative text included prices for each individual product to create a realistic online retail environment. To imitate a realistic shopping situation the environmentally sustainable products were labelled with higher prices (Henninger et al. 2016). In the *LimeSurvey* tool, default answers are defined that pre-select the GFP for each category. In these cases, the radio button showed as already selected and was filled in in blue. For the standard product, the radio button was then displayed in grey (not selected).

Additionally, percentage figures were used as a social norm nudge to describe how many of previous consumers have selected the sustainable product option. The social norm nudge is represented by a green box that was displayed at the bottom of the product image of the GFP. Inside this green box appears the following sentence in German: "X% of the last customers choose this product". For the respective six products, the numbers were randomly assigned to the products, and it was ensured that for each a different percentage figure was displayed. The social norm nudges were divided into three levels: High (90%, 71%), medium (52%, 46%) and low (34%, 20%).

The subsequent questionnaire referred to participants' evaluation of autonomous decision-making aiming to examine whether the participants accepted the nudge as such and did not feel restricted in their free choice (Arvanitis et al. 2020). For this purpose, an adapted version of the autonomy sub-scale by Wachner, Adriaanse, & De Ridder (2020) based on the Basic Psychological Needs in Exercise Scale (BPNES) (Vlachopoulos et al. 2010) was used. In the adapted version, this scale showed a good internal consistency ($\alpha$ = .76) (Wachner et al. 2020). The present study used a five-point Likert scale (from 1 "strongly disagree" to 5 "strongly agree") to measure four items (e.g.: "I feel very strongly that I had the opportunity to influence my decision" or "my decision is highly compatible with my goals and interests").

Furthermore, in order to investigate to which degree the participants identifies with environmental consciousness, the ten-item questionnaire of Brown et al. (1986) is used. The questionnaire has been used in previous studies and demonstrates good internal consistency ($\alpha$ = .82) (Brown et al. 1986). It proves of good applicability as it utilizes a blank space in which the investigated group can be inserted manually by the researchers (E.g.: "I am a person who identifies with the --- group", "I am a person who feels strong ties with the --- group" (Brown et al. 1986)). The extent to which participants identifies with environmentally consciousness is measured by a 5-Likert scale (from 1 "strongly disagree" to 5 "strongly agree"). Eventually, demographic data was collected, focusing on age, gender, profession, and monthly income to collect possible insights what might drive the decision-making process in the context of this study.

# 5 Findings

## 5.1 Descriptive Results of the Sample

In total, 404 participants took part in the study. However, 84 had to be excluded as they did not finish the questionnaire leading to a sample of 320 participants. 77.8% ($N$ = 249) were female while 22.2% ($N$ = 71) were male with a mean age of 32.19 ($SD$ = 11.05). The distribution of participants in the nudging treatment groups was balanced (*default*: $N$ = 87, *social norm*: $N$ = 87, *combination*: $N$ = 87). Only the *control group* showed a reduced number of participants ($N$ = 59).

To investigate the hypotheses, the Cronbach's alpha value of the questionnaires was checked, as well as the normal distribution for the variables. The internal consistency of the scale that assesses the identification of participants with environmental consciousness (Brown et al. 1986) reveals a satisfying internal consistency with Cronbach's alpha for positive affect ($\alpha$ = .77). The participants, on average, rated themselves as more likely environmentally friendly ($M$ = 4.07, $SD$ = 0.58) with a minimum rating of 2.3 and a maximum of 5. Cronbach's alpha of the scale for perceived autonomy of decision-making shows a satisfying internal consistency ($\alpha$ = .73). Descriptive evaluations reveal a mean score of 3.94 ($SD$ = 0.84).

In general, an average of 3.83 ($SD$ = 1.55) GFPs in six buying scenarios were selected by the participants. Inspection the individual groups, the most products were selected within the *social norm* group ($M$ = 4.09, $SD$ = 1.42). An average of 3.93 ($SD$ = 1.58) products were selected in group *default* and 3.75 ($SD$ = 1.57) products in the *control group*. The fewest number of products was selected in the group *combination* ($M$ = 3.53, $SD$ = 1.59).

## 5.2 Analysis of Digital Nudging Strategies

To test hypotheses 1a and 1b which assume the individual use of digital default ($M$ = 3.93, $SD$ = 1.58) and social norm nudges ($M$ = 4.09, $SD$ = 1.42) to enhance GFP choices compared to decision settings in which no nudge ($M$ = 3.75, $SD$ = 1.57) is applied, a one-way ANOVA was calculated. The treatment group is used as the independent variable and GFP selection represents the dependent variable. Checking the normal distribution of the dependent variable GFP selection revealed that data was not normally distributed (Schapiro-Wilk test, $p$ < .001). However, since the test is highly sensitive, by inspecting the Q-Q plot of each group (Appendix) an approximately normal distribution for each group can be assumed. In the absence of variance homogeneity (Levene's test, $p$ = .484), the output of the Welch-Test is considered, which reveals that the number of GFP did not differ significantly between the treatment groups, Welch-Test $F(2,140.35) = 0.93$, $p = .396$. Therefore, the hypotheses 1a and 1b can be rejected.

For the analysis of the second hypothesis, which assumes that all applied nudge strategies increase GFP selection compared to when no or single nudge is applied, an additional one-way ANOVA was performed, taking purchasing decisions of participants in the control group ($M = 3.75$, $SD = 1.57$), *default* ($M = 3.93$, $SD = 1.58$), *social norm* ($M = 4.09$, $SD = 1.42$), *combination* ($M = 3.53$, $SD = 1.59$) into account. Results reveal that the number of selected GFPs did not differ significantly between the four treatment groups, $F(3,316) = 2.132$, $p = .096$. Hypothesis 2, therefore, can be rejected as well.

## 5.3 Explorative Analysis

Further calculations were made to identify possible factors of influence. First, it was determined whether there was a significant difference in the stated environmental consciousness between the different treatment groups which might have impacted the decision to purchase GFP. However, the results indicate that the environmental consciousness did not differ statistically significantly between the treatment groups, $F(3, 316) = 1.44$, $p = .231$. Next, differences in perceived autonomy of decision depending between treatment groups was investigated. In the absence of variance homogeneity (Levene's test, $p = .157$), the output of the Welch-Test is considered, which proves that perceived autonomy did not differ significantly within the treatment groups, Welch-Test $F(3,316) = 1.32$, $p = .268$. The following table represents the average indication for individual treatment groups regarding decision autonomy and identification with environmental consciousness.

To test whether the interaction between participants' environmental consciousness and applied digital nudging strategies significantly predicts GFP selection, a moderation analysis was conducted using PROCESS (Hayes 2018). The overall model was significant, $F(3, 316) = 15.49$, $p < .001$, with a variance resolution of 14.41%. Results failed to find a moderation effect of environmental consciousness on the relationship between treatment groups on GFP selection, $\Delta R^2 = 0.27\%$, $F(1, 316) = 0.92$, $p = .338$, 95% $CI[-0.377, -0.14]$. Since no interaction effect is found, the interaction term was removed from the model, following the recommendations of Hayes (2018). This results in a new main effects model. To calculate the moderation model without including the interaction term, a linear regression was calculated with environmental consciousness and treatment group as independent variables and GFP selection as the dependent variable. Figure 3 illustrates the investigated linear relationship within each treatment group.

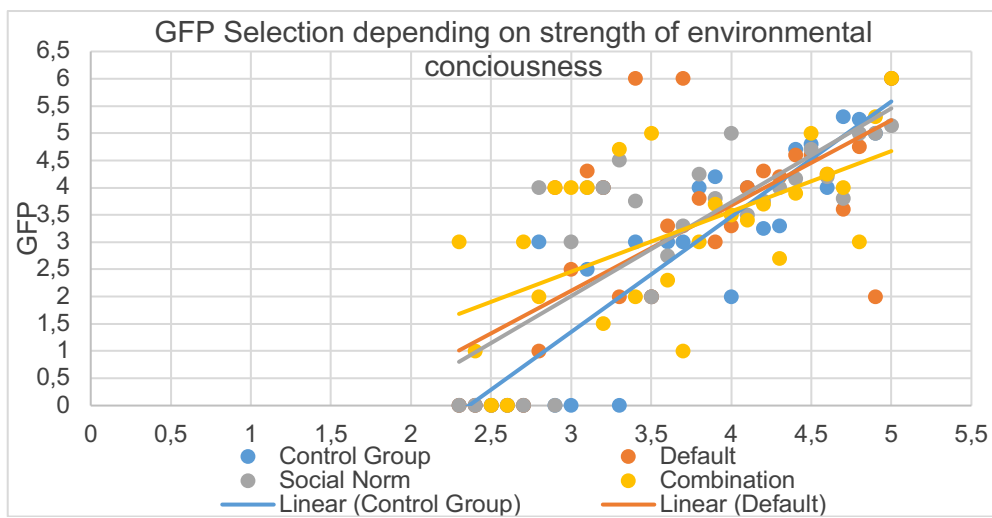

*Figure 3: Distribution showing moderation effect of social identity on GFP selection.*

The new model showed a significant relationship between environmental consciousness, $B = -1.01$, $p < .001$ and GFP selection while no significant relationship was found between nudging strategy, $B = -0.04$, $p = .615$ and GFP selection. The conducted linear regression reveals that if environmental consciousness increases by one unit, GFP selection increases by 1.01. This linear regression was conducted for each treatment group. When participants' environmental consciousness increases by one unit in the *control group*, GFP selection increases the strongest ($B = 1.61$), whereas for *default* ($B = 0.78$) and *social norm* ($B = 0.79$), GFP selection increases minimal. Group *combination* showed an increase of 0.94.

## 6　Discussion

The main goal of the present study was to examine the applicability of nudge strategies to enhance the motivation of participants for GFP purchases. Based on the results of the additional analysis that revealed significant impact of the participants' identification with environmentally sustainability, the proposed research model was adjusted and moderation effects of identification with environmental sustainability was added (Figure 4).

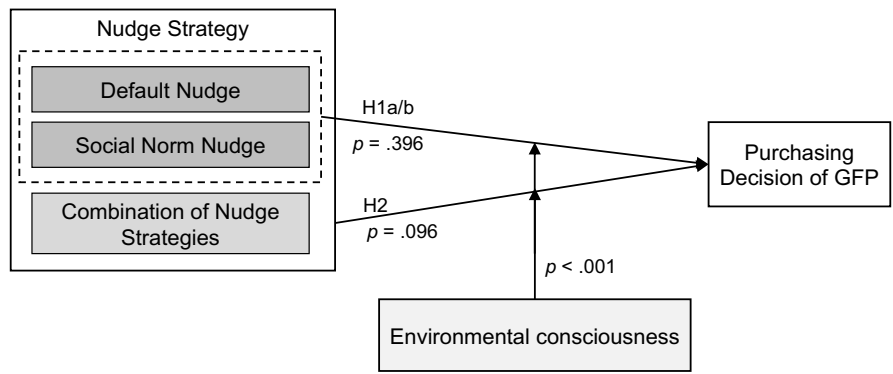

*Figure 4: Adjusted research model including test results.*

### 6.1　Effect of Digital Nudging Strategies on Purchasing Behaviour

This study tested whether the application of digital nudging strategies significantly increases purchasing behaviour of GFP. The analysis of hypotheses did not reveal any statistically significant differences of the purchasing behaviour between the treatment groups and more precisely, that all participants statistically showed similar behaviour. These results contradict previous studies which prove that social norm and default nudge strategies significantly motivate consumer behaviour for purchasing green products (e.g., Aldrovandi et al., 2015; Charlier et al., 2019; Demarque et al., 2015; Ingendahl et al., 2020) and that the combination of digital nudging strategies describes the most effective nudging method (e.g., Charlier et al., 2019; Ingendahl et al., 2020). However, a failure of a digital nudging strategy is not a bad outcome (Osman et al. 2020; Sunstein 2017).

Descriptive GFP selection between *control group*, *default* and *social norm* show that while participants in the *control group* selected the lowest number of GFP, participants in *social* norm selected the highest number of GFP. In combination with the non-significant difference in environmental consciousness, a positive effect on motivation participants for GFP selection by the social norm nudge can cautiously be assumed. In this context, however, the concept of "social defaults" in "uncertain products" could impact the decision-making. Consumers who do not know enough about a product to be evaluated tend to go with social defaults (Huh et al, 2014). Even though, we presented the participants with options similar in quality, consumers who are not particularly fashion-savvy may just go with the social default option rather than the "green-ness" or sustainability factor, simply because they have difficulties to assess the quality outright. Moreover, participants could be 'optimising' for other variables such as the price in certain contexts, although we tried to provide green fashion products with a realistically higher price tag. Nevertheless, inspecting descriptive results leads to the assumption that participants in group *combination*, in which two nudge strategies were combined, chose the lowest number of GFP compared to the other treatment groups. Although a small effect, the difference between groups can lead to the assumption that a *backfire effect* of the combination strategy is present (Osman et al. 2020). The backfire effect describes that the nudging strategy led to changed consumer behaviour, but in the opposite direction than it was intended and that results can be observed as being even lower than in the control condition (Hummel and Maedche 2019; Mols et al. 2015; Osman et al. 2020).

It was found that especially excessive approaches of nudging strategies, including their combination, might lead to possible reactance and perceived compulsion in decision-making and elicit counter-reactions of individuals. In these cases, failures occur as individuals feel limited in their autonomy of decision-making (Osman et al. 2020; Sunstein 2017). However, as the analysis found no statistically significant difference in autonomy perception between groups, it can be assumed that the selected nudging strategies ensured free decision-making (Halpern 2015; Thaler and Sunstein 2003).

### 6.2 Moderation Effect of Environmental Consciousness

Results of the explorative analysis reveal the importance of the individual identification with the topic that is subject to the nudging. First, results found a statistically significant influence of environmental consciousness on the relationship of nudging strategies and GFP selection. This implies that for each nudging strategy participants with the same strength of environmental consciousness were motivated for GFP selection to different degrees. In other words, it can be assumed that one applied nudging strategy is better than another for individuals with the same strength of environmentally consciousness. This result supports previous research that argues some nudging strategies might be more effective in situations in which others might be not (Sunstein 2017).

Investigating the scatter plot shows that for strong identifiers in group *combination*, there is a reduced product selection compared to other treatment groups. Additionally, it can be detected that those weak identifiers selected more products than weak identifiers in other treatment groups. However, it seems that the nudging failure is so strong that it triggers the overall backfire effect, despite the partially increased selection of weak identifiers. This serves as an important result that supports the argument of Mols et al. (2015), which highlights the need to identify factors from a social psychological perspective, to conclude which factors influence the effectiveness of nudges, and to assure sustainable and ethical behavioural change. For instance, referring to the effectiveness of nudging strategies, previous research found that a lower identification with environmental consciousness is associated with a lower purchasing behaviour (Bly et al. 2015; Park and Lin 2020) and that uncertainty towards a certain topic as well as existing conflicting opinions reduces the nudges effectiveness (Venema et al. 2020). Even though participants in the group *combination* indicated the lowest identification, their indication is above the mean of the queried range and therefore rather strong. The results of this study extend the work of Osman et al. (2020) in which nudging failures are analysed and declared into different categories such as "no treatment effect" or "backfiring".

## 7 Conclusion

This study examined the potential of digital nudges to motivate GFP purchasing behaviour and further found a hitherto neglected impact in the context of nudging strategies. Although the results revealed a statistically non-significant difference between the treatment groups in terms of GFP purchasing behaviour, by inspecting descriptive results, a positive influence of the default and social norm nudges on GFP purchases can cautiously be assumed. Explorative results, however, reveal that an overall backfiring effect can be observed. This study comes with limitations. The equal distribution of participants was not ensured as fewer participants were included in the control group than in the nudge groups. Furthermore, the analysis shows an unequal balance of gender since more female participants took part in this study. Although a positive effect can be assumed, the statistically non-significant difference of GFP selection between groups limits the occurring positive influence on GFP selection and a backfiring effect with a positive side effect to an assumption. Further research on the presented issue is relevant to limit the negative environmental impact of online fashion retail.